\documentclass[manuscript,authorversion,nonacm]{acmart}
\AtBeginDocument{%
  \providecommand\BibTeX{{%
    \normalfont B\kern-0.5em{\scshape i\kern-0.25em b}\kern-0.8em\TeX}}}

\copyrightyear{2024}
\acmYear{2024}
\acmDOI{XXXXXXX.XXXXXXX}

\usepackage{multirow}
\usepackage{graphicx}
\usepackage{outlines}
\begin{document}

\title[Reconnecting International Travel: The Personal Infrastructuring Work in Crisis]{Reconnecting An International Travel Network: The Personal Infrastructuring Work of International Travelers in A Multi-facet Crisis}

\author{Yao Lyu}
\email{yml5549@psu.edu}
\orcid{0000-0003-3962-4868}
\author{He Zhang}
\email{hpz5211@psu.edu}
\orcid{0000-0002-8169-1653}
\author{John M. Carroll}
\email{jmc56@psu.edu}
\orcid{0000-0001-5189-337X}
\affiliation{%
  \institution{Pennsylvania State University}
  \city{University Park}
  \state{Pennsylvania}
  \country{USA}
  \postcode{16802}
}

\renewcommand{\shortauthors}{Yao et al.}

\begin{abstract}
In times of crisis, international travel becomes tenuous and anxiety provoking. The crisis informatics and Human-Computer Interaction (HCI) community has paid increasing attention to the use of Information and Communication Technologies (ICTs) in various crisis settings. However, little is known about the travelers' actual experiences in whole trips in crises. In this paper, we bridge the gap by presenting a study on Chinese travelers' encounters in their international journeys to the US during a multifacet crisis and their use of ICTs to overcome difficulties in the journeys. We interviewed 22 Chinese travelers who had successfully come to the US during the crisis. The findings showed how travelers improvised to reconnect the broken international travel infrastructure. We also discuss the findings with the literature on infrastructure, and crisis informatics, and provide design implications for travel authorities and agencies.      
\end{abstract}






\maketitle
\section{Introduction}
International travel invariably brings with it a host of complexities and uncertainties. As travelers journey between countries, they navigate diverse cultures, protocols, and infrastructures. Such challenges can potentially lead to unanticipated delays or changes in plans. This inherent uncertainty intensifies during crises, often resulting in situations that lack clear guidelines or where travelers must quickly adapt to changing regulations~\cite{Hlavka2021}.

From 2020 onwards, the world of international travel underwent transformative disruptions. Primarily, the COVID-19 pandemic, a momentous health crisis, enforced numerous restrictions in public domains. These policies, influenced by public health and political deliberations, curtailed global mobility significantly. By June 2021, the pandemic had culminated in a drop of roughly 4 billion airline passengers, leading to an economic downturn of about 600 billion US dollars, as reported by the International Civil Aviation Organization (ICAO)\cite{AirTransportBureau2021}. Furthermore, as the virus and its subsequent strains propagated globally\cite{Seyfi2020}, governments responded with measures such as lockdowns and travel bans~\cite{Kanno-Youngs2021}. The geopolitical landscape too, exemplified by the Russia-Ukraine tensions starting February 2022, resulted in countries reevaluating their border policies~\cite{ElaineGlusac2022}. 

A closer look at the China-US travel corridor reveals the compounded challenges faced. Health considerations led the US to restrict travelers from numerous regions, including China, to protect its domestic populace~\cite{pro9984}. These directives mainly barred travelers who had visited specific nations in the prior 14 days, excluding certain categories like US citizens and immediate families. Historically, the US has also employed travel restrictions for security reasons. For instance, in 2017, several primarily Muslim countries faced a travel ban due to national security concerns~\cite{pro9645}. Similarly, by June 2020, the US introduced restrictions on select Chinese students and researchers, citing security risks~\cite{pro10043}. These policy changes can have a significant impact on future planned trips, and it is particularly important to have this information in a timely manner in order to reduce travel expenditures. Therefore, this paper focuses on how Information and Communication Technologies (ICTs) can help mitigate this.

In this study, the term "multi-facet crisis" refers to the turbulent global atmosphere after 2020, characterized by disruptions in international travel due to health and geopolitical issues. We narrow our focus to directives from the US government that significantly impacted Chinese travelers. Here, we use "Chinese travelers/passengers" to refer to Chinese citizens, who did not have the privilege of US citizens or Chinese citizens who were spouses, parents, or children of US citizens, during their trips from China to the US. 

Although the severity of the \textit{multi-facet crisis} has diminished (for example, the easing of the pandemic), the nuanced \textit{in situ} experiences of these travelers during this multifaceted crisis remain largely unexplored. This paper fills this gap, delving into the challenges faced by Chinese travelers on their journeys to the US and their reliance on Information and Communication Technologies (ICTs) to mitigate these obstacles.

From the perspective of the \textit{multi-facet crisis}, Chinese travelers had to navigate both health advisories and policy impediments. Balancing the public health warnings against international travel with the intricacies of journeying between two politically tense nations presented numerous socio-political and identity-related challenges. To further understand the influences and challenges, we posed the following research questions:

\textbf{RQs: What were the challenges in Chinese travelers' journeys to the US in the multi-facet crisis and how did they address the challenges?}

We conducted a qualitative interview study to answer the research questions. We recruited 22 Chinese travelers who had successfully come to the US in the crisis. We used narrative interviews \cite{Jovchelovitch2000} to document their experiences in their trips, including the intent of, the preparation for, the actual itinerary of the trips, as well as the challenges in the trips and the practices for addressing the challenges. We then used an inductive thematic analysis \cite{Braun2006b} to interpret the interview data. In the data analysis stage, we observed a salient pattern among participants, which contrasted travel experiences before and after the crisis. Participants missed the smooth experiences before the crisis and then complained about how much work they had to do to address the current problems in the crisis. Such a pattern reminded us of the concept of infrastructure, which refers to a large-scale socio-technical system that provides stable and seamless services \cite{Star1996}, and infrastructuring work, which denotes users' work of repairing the infrastructure when it breaks down \cite{Pipek2009}. Therefore, we used infrastructure and infrastructuring as an analytic lens to comprehend travelers' experiences. 

The analysis resulted in three themes. The first theme is about overcoming restrictions due to travel policies; participants revealed that they encountered travel restrictions due to both public health and geopolitical reasons, and due to the lack of official support, they had to rely on work by themselves or with help from other travelers. The second theme is stitching seams in the itinerary; travelers, in addition to official departments, also had to deal with tourism services like airlines, hotels, and local authorities during stopovers; due to the crisis, tourism services also became unreliable, and travelers had to spend extra effort to make sure the tourism services functioned well in their trips. The third theme pivots around the covid pandemic. Travelers on the long journey also needed to pay attention to the risks posed by coronavirus; they had to protect themselves from being infected and manage their covid test reports on their journeys. We then provided two cases of how travelers used ICTs for whitewashing, a specific strategy to overcome the breakdown of travel infrastructure. We also discussed the findings with the literature on infrastructure as well as crisis informatics and tourism.

Our study not only reflects on Chinese travelers' experiences but also speaks to an ongoing situation where travelers, as individuals, have to deal with international travel problems by themselves when public health emergencies, geopolitical tensions, and even wars occur. In addition, the study also contributes to the literature in three aspects: 1) This paper documents how travelers encountered a multifacet crisis. Specifically, we report the multiple challenges induced by the crisis; we also demonstrate travelers' countermeasures to the challenges. We provide a detailed and in-depth account of travelers and their experiences in terms of rebuilding their international travel networks. 2) Based on participants' experiences, we conceptualize travel infrastructure from a crisis informatics point of view, highlighting the importance of ICT infrastructure and human infrastructure. 3) We foreground the improvisation and collaboration of travelers in crisis, providing a more nuanced interpretation of travelers' infrastructuring work.

\section{Related Work}

\subsection{Travelers in Crises}

Traveling from one place to another is an indispensable part of human life. Travelers have to be concerned about various factors to make their trips safe, pleasant, and convenient, The factors include weather~\cite{10.1145/3098279.3122140}, safety~\cite{10.1145/3123024.3124422}, accessibility~\cite{10.1145/3290605.3300246}, as well as route planning ~\cite{10.1145/3213586.3226217,10.1145/3209219.3213591}. When crises happen, the concerns become even more important. The tourism literature has documented numerous studies on tourists in crises~\cite{Sonmez1999,Booyens2022,Lepp2003}. In this section, we introduce tourism literature on travelers in crises settings.

Researchers have investigated travelers' diverse reactions to crises. Among the studies on travelers' perceptions, many tourist scholars adopt the three constructs of the Theory of Planned Behavior (TPB), namely attitude ("\textit{the degree to which a person has a favorable or unfavorable evaluation or appraisal of the behavior in question}\cite{Ajzen1991}"), perceived behavioral control ("\textit{the perceived ease or difficulty of performing the behavior and it is assumed to reflect past experience as well as anticipated impediments and obstacles}\cite{Ajzen1991}"), and subjective norm ("\textit{the perceived social pressure to perform or not to perform the behavior}\cite{Ajzen1991}"), to comprehend the perceptions in crises~\cite{Liu2021b,Rahmafitria2021,Shin2022,Wang2022tpb}. For instance, Rahmafitria et al.~\cite{Rahmafitria2021} applied TPB to analyze how authorities' physical distancing regulation in the pandemic influenced Indonesian people's travel perception. The results showed that subjective norms, i.e. the social pressure caused by collectivism which requires people to comply with the authorities, had a negative influence on the intention of traveling. In addition, scholars also explore other aspects of travelers' perceptions, like trust and fear~\cite{Secilmis2021,Zheng2022,Zheng2021,Rather2021}. Shin et al.~\cite{Shin2022} found that travelers' perceptions were affected by their trust in governments, local authorities, and other passengers; Zheng et al.~\cite{Zheng2021} reported that the covid-19 pandemic's severity caused travelers' "\textit{travel fear}," including being frightened, nervous, and anxious. Some researchers also pay attention to travelers' decision-making in crises. Generally speaking, travelers have to take more effort when traveling in crises, like maintaining mental health~\cite{Lenggogeni2021}, taking more protection~\cite{Zheng2021}, and so on. For example, Ren et al.~\cite{Ren2022} introduced a case where passengers also needed to pay attention to "\textit{remote effect}" when making travel plans. The remote effect is the influence of the pandemic at the national level on one particular city: although the city might not have many cases currently if the country's other cities' situations are serious, passengers from those cities could still bring the virus into the city.

Among the research on travelers' reactions to crises, one group of researchers pays attention to the importance of travelers' information consumption~\cite{Mizrachi2020,Liu2022}. On the one hand, informative channels, like news websites, in crises are important sources for travelers to make sense of crises. Media coverage presents updated situations of crises~\cite{Chemli2022} and hence influences travelers' decision-making~\cite{Qiao2022}. Travelers need to get enough knowledge before they can make proper strategies~\cite{Rahmafitria2021}. Qiao et al.~\cite{Qiao2022} conducted a survey among Korean people who planned to have international trips during the pandemic. The survey results indicated that media reports on the countermeasures taken by local governments and residents could shape travelers' plans of self-protection. Furthermore, media platforms also play an important role in framing the images of destinations. The framing could significantly influence travelers' behaviors, too. Mayer et al.~\cite{Mayer2021} demonstrated an investigation on how media coverage reported on a super-spreader destination (Ischgl, Austria) in the pandemic and how the reports affected the image of the destination. The results showed that the media's stigmatization of the destination, like calling it "Wuhan of Europe" or "Virus Slingshot", tremendously harmed the image of the destination. The damage to the destination image further hinders travelers' intention of going there~\cite{Lu2021z}. On the other hand, information platforms like social media allow travelers to exchange information with others. Travelers can share their experiences and give helpful recommendations to those who are making travel plans and have not yet started their journeys~\cite{10.1080/13683500.2020.1752632,Rather2021}. Yu and Zhou~\cite{10.1080/13683500.2020.1752632} reviewed more than 10k covid-related comments from Trip Advisor, a platform for travelers to share reviews. The findings pointed out that travelers shared information about the quality of tourism services, local authorities' efforts against the pandemic, media coverage reliability, and discrimination against travelers. Secilmis and colleagues~\cite{Secilmis2021} studied the effect of travel influencers, one type of professional travelers who share their stories with and make recommendations to their audience on social media platforms, on ordinary travelers' perceptions and intentions in the pandemic. The outcome of the study pointed out that the influencers' expertise and experience could help other travelers develop strategies for their own trips.   

In this study, we join the literature on travelers' experiences in a crisis setting. We specifically elucidate the ICT-supported practices by travelers who were restricted by a public health crisis as well as political tensions. We use a qualitative approach to showcase travelers' contextual perceptions and situated interactions when they faced multiple threats.  

\subsection{Crisis Informatics and Travel}
Our research project investigates ICT-mediated work conducted by international travelers in a multi-facet crisis. Therefore, we also build our study on the current literature on crisis informatics. Crisis informatics is an interdisciplinary field where researchers study users' leverage of ICTs in crisis settings. The concept of crisis covers widely, and it generally can be categorized in to three types~\cite{Shaluf2007}: 1) natural disasters caused by nature environment, like wildfires~\cite{Shklovski2008}, landslides~\cite{10.1145/2998181.2998290}, earthquakes~\cite{10.1145/3274430}, and floods~\cite{10.1145/1718918.1718965}; 2) human-made incidents triggered by social, economic, and/or political reasons, like gun violence~\cite{10.1145/3134727}, wars~\cite{10.1145/3491102.3502126}, diasporas~\cite{10.1145/3432946}, and identity crisis~\cite{10.1145/3491102.3517600}; 3) and hybrid ones resulted from both natural and human factors, like infectious disease outbreak~\cite{10.1145/3491102.3517572}. In the context of the current study, the multi-facet crisis is a hybrid crisis where international travel is restrained by both human and natural crises. Travelers have to face the risk of the covid-19 pandemic as well as the travel ban at the same time. We review crisis informatics literature relevant to information seeking for traveling in the following section. 

The crisis informatics community has been paying constant attention to the affordance of various types of ICTs in crises, like contact tracing systems~\cite{10.1145/3491102.3517572} and mobile phones~\cite{10.1145/3173574.3173960}. For instance, Lyu and Carroll~\cite{10.1145/3491102.3517572} presented a study on Chinese residents' use of contact tracing apps in the covid-19 pandemic; The study showed that Chinese residents used contact tracing apps to not only navigate themselves in an infectious disease outbreak but also guide their social lives. Among the research on ICT affordance, one of the most popular topics is social media platforms. Social media, because of its feature of allowing users to exchange information in real-time, has the advantage of providing the most updated information when crises happen. Information on social media sometimes could even be more updated and accurate than official crisis responders~\cite{10.1145/2998181.2998290}. Therefore, information dissemination and seeking on social media is of great significance to crisis response. Haq et al.~\cite{10.1145/3543434.3543642} discuss the issue of governments relying on the versatility of social media to communicate with citizens, focusing on citizen engagement and response to outbreak information. Yang et al.~\cite{10.1145/3491102.3517591} investigated online discourses on Weibo, one of the largest micro-blogging platforms in China, during the pandemic. The discourses, which were published by either covid patients or their caregivers, were about getting urgent healthcare services or medicine. The patients and caregivers used various functions like sending private messages to or mentioning ("@") users who might provide help, using hashtags related to covid-related assistance, and providing personal information to be better identified by responders. In addition to studies focusing on one particular social media, some scholars also show interest in the cooperation across different social media platforms. Norris et al.~\cite{10.1145/3359200,10.1145/3512955} showcased the coordination among the members of a digital humanitarian group, which is globally distributed, after the 2017 Hurricane Maria landfall in Puerto Rico. The team members used various social media platforms like Facebook and Twitter, as well as instant messaging services like Slack, for \textit{information triage}, which refers to "\textit{the process of sorting through (the possibly numerous) relevant materials, and organizing them to meet the needs of the task at hand}~\cite{Marshall1997}."

Studies on social media in crises cover multiple types of users, including authorities and ordinary users. Research of authorities on social media looks for better approaches to provide more timely, accurate, and reliable information to the public. For instance, Li et al.~\cite{10.1145/3449209} looked for better authority response to help-seeking posts on social media. They investigated the linguist features of posts that could be better identified by authorities. Results showed that authority users had a higher possibility to respond to posts with certain features like mentioning specific disasters or expressing sentiments. That said, researchers are paying growing attention to ordinary people's work in crises when the official response is perceived as unreliable or inaccessible~\cite{10.1145/2858036.2858109}. Most ordinary people are neither knowledgeable nor experienced in dealing with a crisis. They also have limited time to practice themselves to become knowledgeable or experienced when crises take place. Therefore, they oftentimes turn to intuition or improvisation in crises. For instance, Kou et al.~\cite{10.1145/3134696} introduced a case where Reddit users collectively made sense of and composed conspiracy theories about the Zika virus outbreak; the primary reason for the sense-making themselves was distrust of or dissatisfaction with the information provided by authorities. Despite the negative consequences, collaboration among ordinary people, however, can be beneficial, too. Das and Semaan~\cite{10.1145/3491102.3517600} reported the collaboration among South Asian Bengali users of BnQuora, a social Q\&A platform in the Bengali language, to cope with the identity crisis due to the colonialism history. The collaboration helped heal users from the colonial trauma. Such studies showcase the importance of understanding grassroots work in the face of crises and foreground the significance of supporting those beneficial work in crises in ICT design.

A few crisis informatics researchers have explored topics related to travel, with a specific interest in how people use social media for their journeys in crises~\cite{10.1145/3025453.3025891,10.1145/3173574.3173788}. A most representative example is Gui et al.'s work~\cite{10.1145/3025453.3025891}. They investigated discourses among users, who planned to travel in the Zika virus outbreak, on three social media sites: Reddit, Trip Advisor, and Baby Center. The findings revealed that users found the information from authorities was problematic and therefore they faced tremendous uncertainty while making travel plans. They had to look for more correct, accurate, and personalized information by themselves. In terms of reducing uncertainty, the study emphasized the importance of the information provided by local people who live in the destinations, the reasoning based on the currently available information, and the calculation of costs and benefits. However, little has been explored on the use of ICTs like social media during their trips.

That said, the investigation of travelers' use of ICT is mainly about the preparation and the intention before their journeys. Little is known about travelers' actual experiences during the whole trips in crises, including the real-time challenges caused by the crises and the ICT-supported approaches to addressing those challenges. In this paper, we bridge the gap by presenting a study on Chinese travelers' encountered in their international journeys to the US and their use of ICTs to overcome difficulties in the journeys.

\subsection{Infrastructuring Work in Crises}

In our study, all participants mentioned that they experienced the tremendous influence of the crisis. Specifically, they describe the influence as a situation where all the seamless travel services suddenly disappeared and travelers had to overcome it by themselves, like manually arranging schedules of connecting flights, and carefully reviewing information about hotels before booking. The situation showcases the breakdown of a foundation that travelers rely on and travelers' scaffolding work to fix the breakdown. It further reminds us of the concept of infrastructure and infrastructuring work, which refers to a socio-technical foundation that supports certain human activities and the work of repairing the foundation by users after its failure~\cite{Star1996,Pipek2009}. In this section, we elaborate on infrastructure studies related to our work.

In field of Human-Computer Interaction (HCI) and Computer-Supported Cooperative Work (CSCW), infrastructure refers to large and distributed socio-technical systems~\cite{Star1996}. It includes not only devices or systems but also the users and users' work around it. According to Star and Ruhleder~\cite{Star1996}, infrastructure has eight characteristics: 1) infrastructure is embedded into its environment; 2) infrastructure is transparent when at work; 3) infrastructure can reach beyond a single event or site; 4) new participants learn infrastructure through becoming a member of it; 5) infrastructure shapes and is shaped by conventions; 6) infrastructure is plugged into other infrastructures in a standardized manner; 7) infrastructure is built on installed bases; 8) infrastructure becomes visible upon breakdown. Infrastructure researchers emphasize the "\textit{when}" instead of the "\textit{what}" of an infrastructure. Infrastructure emerges "\textit{when local practices are afforded by a larger-scale technology, which can then be used in a natural, ready-to-hand fashion}~\cite{Star1996}." In other words, infrastructure does not have a fixed structure, depending on what currently available resources users can leverage and what tasks they want to accomplish. 

Therefore, scholars have conceptualized various infrastructures according to its constitution, such as \textit{human infrastructure}~\cite{Lee2006} (the organization of human labor that supports certain activities). Recently, scholars are paying attention to conceptualizing infrastructures based on specific purposes they serve. Britton et al.~\cite{Britton2019} introduced "\textit{critical infrastructure}" for new mothers who are experiencing significant life transitions. Critical infrastructure, in this context, refers to "\textit{digital and non-digital spaces through which people reflect on their own identities while also critiquing existing systems}~\cite{Britton2019}." The digital and non-digital spaces include Facebook groups and local communities for mothers. New mothers used such infrastructure for "\textit{simultaneously renegotiating the ideals of motherhood while also critiquing the gendered norms that permeate their social worlds}~\cite{Britton2019}." The literature also covered infrastructure based on social media. Dailey and Starbird used "\textit{crisis infrastructure}~\cite{10.1145/2998181.2998290}" to refer to the combination of multiple information infrastructures, including various social media platforms like Twitter and Facebook, for the communication between crisis responders, journalists, and residents in the 2014 Oso Landslide. In our study, international travel needs the support of various ICTs, like flight/hotel booking websites or travel advice platforms, and human labor, such as customer services. They together form a foundation that supports international travel, and without it, international travel could be significantly hindered. Therefore, the ICTs and human labor can be considered as travel infrastructure.

Infrastructure does not work perfectly all the time. For example, Vertesi~\cite{Vertesi2014} used "\textit{seams}" to describe infrastructural problems where "each system lies in messy and even un-articulated local overlap with other systems~\cite{Vertesi2014}." When infrastructure breaks down, it causes inconvenience or even trouble to users. Gui and Chen~\cite{10.1145/3290605.3300688} reported the breakdowns of healthcare infrastructure in the US, including healthcare workers' mistakes, healthcare departments' mis- or non-alignments, and the constraints embedded in the healthcare infrastructure. Patients and caregivers encountered huge challenges due to the breakdowns and they had to work by themselves to fix the breakdowns. The fixing work conducted by users is called "\textit{infrastructuring work}~\cite{Pipek2009}." Infrastructuring work is the activities that make an infrastructure work~\cite{Pipek2009}. It is different from "\textit{notions of design that only refer to professional, or to put it better, professionalized design activities}~\cite{Pipek2009}." Instead, it focuses on users' contributions to infrastructure design. Infrastructuring work has many types, depending on the specific context. For instance, Rajapakse et al.~\cite{10.1145/3196709.3196749} dubbed "\textit{personal infrastructuring}" to denote the work by people with disabilities and/or their caregivers "\textit{to understand a variety of services and infrastructures from the point of view of their personal needs, and also to draw together an infrastructure that supports themselves}~\cite{10.1145/3196709.3196749}." In addition to categories, researchers also delve into how users look for feasible solutions. Lyle et al.~\cite{10.1145/3173574.3174026} introduced a type of infrastructuring work that consisted of \textit{strategies} and \textit{tactics}, where strategies refer to "\textit{actions which, thanks to the establishment of a place of power (the property of a proper), elaborate theoretical places (systems and totalizing discourses) capable of articulating an ensemble of physical places in which forces are distributed}~\cite{DeCerteau1894}" and tactics refer to "\textit{calculated action(s) determined by the absence of a proper locus. No delimitation of an exteriority, then, provides it with the condition necessary for autonomy}~\cite{DeCerteau1894}." Some researchers also attribute infrastructuring work to certain qualities. Erickson and Jarrahi~\cite{10.1145/2818048.2820015} used "\textit{infrastructural competence}" to denote the ability to be "\textit{able to recognize where infrastructural seams may have generative, rather than exclusionary, properties and then to draw upon this sociotechnical insight to fashion and implement an infrastructural strategy to achieve the desired goal}~\cite{10.1145/2818048.2820015}."

Specifically, infrastructure researchers also pay attention to infrastructuring work in crises. The literature has documented various infrastructure breakdowns in crises. For example, Semaan and Mark~\cite{10.1145/2063231.2063235} elucidated the damage to infrastructures in Iraq due to the Second Gulf War. The broken infrastructures included transportation infrastructure (e.g. unsafe roads due to military activities), education infrastructure (e.g. destroyed campus), and information infrastructure (e.g. misinformation manipulated by local authorities). In crises, infrastructure failures cause significant troubles to ordinary citizens and oftentimes the citizens cannot get official assistance timely. They have to conduct infrastructuring work to fix the failures so that their lives can come back to normal. Researchers have investigated infrastructuring work in such critical situations. Some of them focus on citizens' individual ability in the face of crises. Semaan et al.~\cite{10.1145/3359175} researched infrastructuring work by citizens who were suffering from the ongoing disruptions in their daily lives. The research emphasized the importance of resilience, which refers to "\textit{the ability of a technical or human system to 'bounce back' from threat or vulnerability}." In addition to individual quality, some researchers also underscore the significance of social support at the community level. In a most recent study, Hussain et al.~\cite{10.1145/3392561.3394640} introduced a group of Rohingya refugees, who were forced to leave their homeland and reside in Bangladesh camps, due to historical and political reasons. The refugees had to conduct infrastructuring work not only to build their physical homes but also to connect their social ties with each other. The authors identified  \textit{communal hope} among refugees, which "\textit{focuses on a practical and feasible future that is rooted in the morality of a community}~\cite{10.1145/3392561.3394640}."

In our study, we apply the lens of infrastructure and infrastructure to conceptualize the travel infrastructure and to comprehend the work conducted by travelers to fix the infrastructure when it broke down in a crisis. We argue that travelers' work that fixed the travel infrastructure is one type of infrastructuring work which underlines the significance of travelers' creativity in the face of infrastructure breakdowns. We also emphasize the importance of several key components of travel infrastructure in crisis.

\section{Methods}
We aim at understanding international travelers' lived experiences in the face of a multi-facet crisis in this project. To address the research question, we conduct qualitative research consisting of an interview study and thematic analysis. In this section, we first introduce the data collection methods, including how we design the interview protocol and how we recruit participants. Then we demonstrate the data analysis methods applied in the study, particularly in analyzing the interview transcripts using thematic analysis. This study has obtained IRB approval from the authors' institution.  

\subsection{Data Collection}

\subsubsection{\textbf{Interview Study Design}} 
We chose narrative interviews as our data collection method. Jovchelovitch and Bauer \cite{Jovchelovitch2000} stated that "\textit{...narrative interview envisages a setting that encourages and stimulates an interviewee to tell a story about some significant event in their life and social context}." Narrative interviews are particularly useful for studies that investigate participants' experiences of specific events under the influence of their "\textit{life histories and socio-historical contexts} \cite{Jovchelovitch2000}." Previous HCI researchers have used narrative interviews to investigate participants' experiences with ICT during events such as seeking healthcare services \cite{10.1145/3290605.3300688,Gui2018} or transitioning to adulthood \cite{Sawhney2018}. In our study, the specific event, life history, and social-historical contexts referred to travelers' journey from China to the US, their social backgrounds, and US's travel ban on Chinese travelers during the covid-19 pandemic, respectively.

In the interviews, we prompted participants to revisit their entire journey - from the initial intention to travel to the US to their eventual settlement in the country. Broadly, participants' memories revolved around themes such as "the purpose of the journey," "pre-journey preparations," "travel timelines and routes," "actual travel experiences," "use of ICTs during the journey," and "contrasts between this journey and previous ones." However, it's worth noting that some interviewees described markedly different experiences, especially concerning flight bookings, schedule planning, and the procurement of VISAs and COVID-related documentation. To comprehensively capture these unique, situated experiences, we delved deeper, posing follow-up questions whenever travelers highlighted specific events or incidents of particular interest to our study.

Considering the situation of the pandemic and the difficulty of arranging in-person meetings with all the participants, who were residing in different states, the interviews were done remotely, either through WeChat, Zoom, or phone calls. We recorded the interviews using electronic devices with the permission of the interviewees. After we finished each interview, we transcribed the audio documents into text transcripts. All the interviews were conducted by the first author in Mandarin. We translated the quotes from transcripts into English word by word. For readability purposes, we also edited the quotes, including deleting utterances and inserting context information, to make the quotes clear and complete. We acknowledge that, even though we try our best to capture the meanings of our participants' quotes, there might be a loss of nuance due to the translation.

\subsubsection{\textbf{Participants Recruitment}}
For this study, we are interested in adult Chinese travelers who have successfully come to the US during the travel ban period. We recruited participants from three approaches. The first author is a Chinese international student who is residing in the US. He contacted Chinese students traveling to the US in the crisis from his social network. Using the social network, he also got access to WeChat groups where Chinese travelers produced, shared, and exchanged information about traveling to the US. He sent flyers about the current study and recruited travelers who have already arrived in the US from the WeChat group. He also asked interviewees to recommend potential participants after each interview (i.e. snowball sampling). The recruitment stopped when the study reached the point of theoretical saturation \cite{Bowen2008}.   

In total, we have 22 participants (See Table \ref{table:demo}). The interviews lasted from 26 to 65 minutes (avg. = 45.6), and we compensate interviewees 10 USD per half an hour for their time in the interviews. When interviewing participants, we asked about their demographics so that we can have an overall understanding of our study population. The interviewees included 13 females and 9 males (self-reported gender); and the age of participants ranged from 21 to 60 (avg. = 32.4). Among the interviewees, one of them graduated from high school, 11 held bachelor's degrees, 9 had master's degrees, and one had a doctorate. Most of them (15 out of 22) were students and came to the US for educational purposes (e.g. starting university, coming back to university, or visiting scholar); another big group of them (6 out of 22) were for family reasons (visiting children's family or looking after kids); one of them was here for expatriate business. We also documented the itineraries of travelers, and most of them (20 out of 22) have been to at least 1 stopover in their journeys.

\begin{figure}[htp]
    \centering
    \includegraphics[width=15cm]{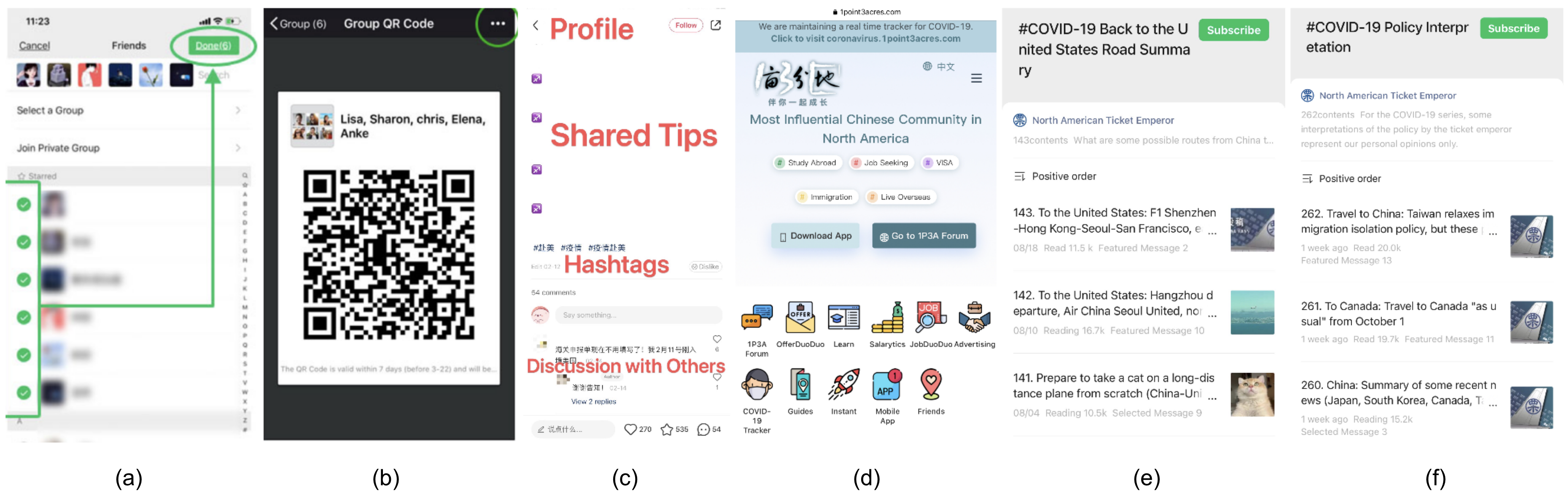}
    \caption{Interfaces of WeChat Group (a,b), Xiaohongshu (c), 1Point3Acres (d), and AirTicketNA (e,f).}
    \label{fig:ICT}
\end{figure}

In addition to the information mentioned above, we also asked participants what ICTs they used for their journeys. All participants used the WeChat Group to get information directly from other travelers. WeChat Group is a function of WeChat, the most popular communication tool in China, that allows users to chat in a group; users can form WeChat Groups by inviting people from their friend lists \cite{WeChat2022a} or providing QR codes \cite{WeChat2022} so that strangers can scan and join. Eight participants got information from Xiaohongshu \cite{Xiaohongshu2022,Alex2022}, a social media platform for purchasing, reviewing, and sharing products, especially items bought by surrogate shoppers from other countries. In addition, eight travelers sought advice from 1Point3Acres.com \cite{1Point3Acres2022}, an online forum for Chinese students and immigrants in North America. Lastly, four participants told us they benefited from the information shared by AirTicketNA \cite{AIRTICKETNA2022}, one organization that specifically shares information about flights and tickets of airlines between China and North American countries. The interfaces of such ICTs are presented in Figure \ref{fig:ICT}. More details about how travelers used the ICTs are elaborated on in the findings section.

\subsection{Data Analysis}

We took a thematic analysis \cite{Braun2006b} approach to comprehend the data collected from the interview study. Thematic analysis is "\textit{a method for identifying, analyzing and reporting patterns (themes) within data... it minimally organizes and describes your data set in (rich) detail... (and) further... interprets various aspects of the research topic} \cite{Braun2006b}." We started the analysis from chronologically organizing interviewees' narratives of their travel experiences. Several key events emerged after the organization, including "initiating the intent to come to the US", "researching the background of traveling to the US", "making sense of international travel policies and local pandemic situation", "booking flights and hotels", "obtaining required documents for travel", "handling unexpected issues during the journey." Specifically, we extracted quotes that describe travelers' use of ICTs in such experiences, including what challenges they faced, what technology they used, and how they, either collectively or individually, used the ICTs to overcome the challenges. We also took travelers' comments, feelings, and other perceptions of such events into consideration. We used these quotes as the units of analysis (codes) because they represented travelers' meaningful, creative work in the face of an uncertain environment for traveling internationally.

After coding all the transcripts, we got 138 codes. We then took a back-and-forth approach to group/regroup the codes so that they could form identical and meaningful themes. In the process, we noticed that all travelers underscored one specific type of experience: their international travel experiences before the pandemic were smooth, so they took it for granted; but during the pandemic, international travel became significantly difficult and the assumed approaches were not working; especially, the official support was missing; therefore, in the face of such unexpected challenges, travelers had to work by themselves to fix them (e.g. T03 in \ref{T03}). This phenomenon reminded us of the concept of infrastructure, which refers to a socio-technical foundation that provides stable and seamless services \cite{Star1996}, and infrastructuring work, which denotes users' work of repairing the infrastructure when it breaks down \cite{Pipek2009}. Infrastructure and infrastructuring work has been applied in numerous HCI studies that investigate users' work to address the breakdown of socio-technical infrastructures \cite{10.1145/3491102.3501949,10.1145/3491102.3502038}. Therefore, we used infrastructure and infrastructuring as an analytic lens to comprehend travelers' experiences, organize codes, and generate themes. Finally, we got three themes from the thematic analysis. We give a more detailed presentation of each theme in the following section. In addition, we also provide a presentation of a case study that specifically focuses on the travelers' ICT-supported practices in the crisis.

\begin{table}
    \centering
    \caption{Demographics of the Participants}
    \resizebox{\columnwidth}{!}{
    \begin{tabular}{ccccccccc}
        \hline
        \textbf{\#} & \textbf{Age} & \textbf{Gen.} & \textbf{Edu.} & \textbf{Vocation} & \textbf{Time} & \textbf{Purpose} & \textbf{Stop} & \textbf{ICT Used} \\ 
        \hline
        T01 & 57 & F & Bac. & Retired    & 09/20 & Visit Family        & 2 & WeChat Group                           \\
        T02 & 46 & F & Bac. & Accountant & 10/20 & Look After Child    & 2 & WeChat Group, Xiaohongshu              \\
        T03 & 57 & F & Bac. & Teacher    & 08/20 & Visit Family        & 2 & WeChat Group                    \\
        T04 & 24 & F & Mas. & Student    & 10/20 & Back to University  & 2 & WeChat Group, 1Point3Acres              \\
        T05 & 57 & F & Bac. & Medicine   & 09/20 & Visit Family        & 2 & WeChat Group                           \\
        T06 & 25 & M & Doc. & Student    & 12/20 & Back to University  & 2 & WeChat Group, 1Point3Acres              \\
        T07 & 22 & M & Bac. & Student    & 01/21 & Start University    & 2 & WeChat Group, 1Point3Acres              \\
        T08 & 23 & M & Bac. & Student    & 01/21 & Start University    & 2 & WeChat Group                           \\
        T09 & 25 & F & Mas. & Student    & 08/21 & Start University    & 2 & WeChat Group                           \\
        T10 & 24 & M & Mas. & Student    & 08/21 & Start University    & 1 & WeChat Group, Xiaohongshu, AirTicketNA \\
        T11 & 24 & M & Mas. & Student    & 08/21 & Start University    & 2 & WeChat Group, 1Point3Acres              \\
        T12 & 21 & F & Bac. & Student    & 07/21 & Back to University  & 0 & WeChat Group, Xiaohongshu, AirTicketNA \\
        T13 & 25 & F & Mas. & Student    & 08/21 & Start University    & 2 & WeChat Group, Xiaohongshu, AirTicketNA \\
        T14 & 24 & F & Mas. & Student    & 08/21 & Start University    & 2 & WeChat Group, Xiaohongshu, 1Point3Acres \\
        T15 & 22 & M & Bac. & Student    & 08/21 & Start University    & 2 & WeChat Group, 1Point3Acres              \\
        T16 & 40 & F & Mas. & Manager    & 08/21 & Expatriate Business & 1 & WeChat Group           \\
        T17 & 25 & M & Mas. & Student    & 08/21 & Start University    & 2 & WeChat Group, Xiaohongshu, 1Point3Acres \\
        T18 & 60 & F & H.S. & Worker     & 12/21 & Visit Family        & 1 & WeChat Group                           \\
        T19 & 28 & M & Doc. & Student    & 12/21 & Visiting Scholar    & 1 & WeChat Group, 1Point3Acres              \\
        T20 & 27 & F & Bac. & Unemployed & 01/22 & Visit Family        & 1 & WeChat Group, 1Point3Acres              \\
        T21 & 25 & M & Bac. & Student    & 01/22 & Back to University  & 0 & WeChat Group, Xiaohongshu, 1Point3Acres \\
        T22 & 27 & F & Mas. & Student & 03/22 & Back to University & 2 & WeChat Group, Xiaohongshu, AirTicketNA \\
        \hline       
    \end{tabular}
    }
    \label{table:demo}
\end{table}

\section{Thematic Analysis: Travelers' Experiences in a Multifacet Crisis}
In this section, we will present three themes that describe the challenges travelers faced and the countermeasures they took. The first theme is about overcoming restrictions due to travel policies; participants revealed that they encountered travel restrictions due to both public health and geopolitical reasons, and due to the lack of official support, they had to rely on work by themselves or with help from other travelers. The second theme is stitching seams in the itinerary; travelers, in addition to official departments, also had to deal with tourism services like airlines, hotels, and local authorities during stopovers; due to the crisis, tourism services also became unreliable, and travelers had to spend extra effort to make sure the tourism services functioned well in their trips. The third theme pivots around the covid pandemic. Travelers on the long journey also needed to pay attention to the risks posed by coronavirus; they had to protect themselves from being infected and manage their covid test reports on their journeys. 

\subsection{Overcoming Restrictions of Travel Policies}

The first theme is about travel restrictions caused by policies. In this section, we introduce the policy restrictions that constrained travelers' experiences and the countermeasures travelers improvised. The policy restrictions were of three types: public health (i.e. travel restrictions due to covid), geopolitics (i.e. bilateral travel bans between China and the US), and institutions (e.g. university departments' requirements). Based on participants' feedback, the policies were not only restrictive but also ambiguous. The ambiguity came from the uncertainty in dealing with officials in a foreign country and got worsened by the pandemic. Travelers had to come up with solutions through improvisation or collaboration with other travelers. Travelers viewed policy restrictions from the standpoint that the US would not allow them unless they looked "harmless", from either public health or a geopolitical perspective. Therefore, participants worked hard to get approval from a foreign country/department. 

\subsubsection{\textbf{Coping with Public Health Regulations}}
Most participants (17 out of 22) reported that their travel to the US was hindered by public health regulations. Starting from February 2020, the US government provided a list of countries, including China, and any person who was in any of these countries within the immediate past 14 days could not enter the US. This decision presented huge challenges to Chinese travelers who had plans to go to the US.

For instance, T05 was planning for a trip to the US for a family reunion; her daughter was living in the US and expected to deliver her baby in August 2020. However, T05's plan was disrupted due to the virus outbreak and the travel restrictions: "\textit{We did not know any way to come, so we postponed our trip. Then we heard from our friends that there's a policy that we can go from a third country, take a 14-day-quarantine, and get us 'whitewashed.' So we just came without hesitation.}" The third country option, also known as "whitewashing," was a countermeasure to the traveler restriction: travelers went to a country which was not on the travel ban list, and stayed there for 14 days. The term "whitewash" also suggests travelers' feelings; they had to appear "cleaned" for acceptance by the US authorities. More details about the "whitewashing" will be introduced in the case study section.

\subsubsection{\textbf{Negotiating with Geopolitical Tensions}}
In addition to covid-centered travel restrictions, Chinese travelers also suffered from travel bans due to geopolitical reasons. As mentioned in the background section, the US government decided to reject the entry, decline the VISA application, and revoke the current VISA of Chinese students who were affiliated with certain Chinese institutions. This travel ban significantly influenced Chinese students' journeys, because, without VISA, students would not have the legal identity to enter the US for their studies. Many participants who were students (15 out of 22) reported their (or their friends') struggles. For instance, from a WeChat Group discussion, T08 heard about one Chinese student's VISA was revoked right on her way back from China to the US: "\textit{Because her undergrad school in China was in the sanction list... anyway, she was rejected when she tried to take the transition flight from Tokyo to the US... What a poor girl. She already studied there for years, and her dog, house, and everything were still in the US.}" The student's experience suggested a breakdown of the travel infrastructure where Chinese students' legal identities were protected well. The cost of being a victim was tremendous: not only her legal identity but also her life in the US was snuffed out. 

Such heavy punishment terrified participants and pushed them to adapt to the new condition. Participants reported specific problems in the new condition, including applying for VISA and using a currently valid VISA, and how they worked to solve the problems. T19 showed how he dealt with the dilemma when he applied for VISA: "\textit{I was applying for funding from the Chinese Scholarship Council (CSC). But finally, I didn't take that money, because of the VISA thing. If you check 1point3acres.com, they reported after they got the funding from CSC and from STEM, they could be rejected immediately. It doesn't mean I can't take the money, I just did not take the risk. Finally, I came here at my own expense.}"

\subsubsection{\textbf{Satisfying Institutional Requirements}}
Other than the international level restrictions, some student travelers (4 out of 23) also reported the requirements from their institutions, like asking students to come to the US before semesters began and attend in-person classes. And the consequence of failing to satisfy the requirement was fatal. T09 came to the US in 2021, but she actually got an offer from a US university in 2020. But she could not apply for a US VISA then because of the closure of consulates in China. So she had to give up and reapply in the next year: "\textit{I got an announcement from the US university that gave me offer, saying if I can't get into US before the Fall semester started, then I can't get the funding, because I wouldn't have any American bank account or SSN. But without funding I would not be able to finish my PhD in the US. So, they suggested I reapply next year. Everything happened so fast, and it seems like there was no room for negotiation.}" The requirement from institutions was not the primary reason that hindered students' international journeys, yet they pushed students to manage their journeys within a limited time.

\subsection{Stitching Seams in Itinerary}
The second theme pivots around the seams of travelers' itineraries. Itinerary here refers to travelers' entire journey, including the experiences of both traveling from one place to another with the help of transportation services (trip segments) and staying or resting in a place between their trips (stopovers). Unlike travel restrictions implemented by public health authorities, governments, or institutions in the US, seams of the journeys were caused by the failure of tourism industry-supported services. As emphasized by participants, the travel experiences, including booking tickets from airlines and hotels, taking flights, and transferring or laying over between trips, were seamless before the crisis occurred. However, it became seamful due to the failure of cooperation among different tourism departments, like airlines, hotels, and other services in local cities where travelers departed, laid over or arrived.

\subsubsection{\textbf{Bridging Gaps Between Trip Segments}}
\label{T03}

The first type of seams took place in travelers' trips, especially when travelers were preparing for or moving to the next destinations. Specifically, most of the participants (20 out of 22) reported gaps in the services like airline or local transportation. Some participants complained about the cancellation of flights before the trips. For instance, T21'said, "\textit{a lot of pilots were infected, so the airline canceled all the flights to the US, including mine.}" The massive cancellation of flights was sudden, resulting from the failure of protection on airline staff against the pandemic. Such experience was consistent with prior work that found travelers' trips were disrupted by the sudden cancellation of flights in the pandemic \cite{Zha2022,Liu2021a}. 

In addition to unexpected cancellations, travelers also reported problems with transitioning in airports. T03, a 57-year-old passenger who could not speak any foreign language, showed how she struggled to ask local staff to find her boarding gate in a Korean airport for transition: "\textit{It was in the pandemic, most service windows were closed, and I don't know English or Korean...Eventually, I downloaded Youdao Translator (a Google Translator-like app). I type in Chinese and it translates it into English. We "talked" in this way, and we use some body language, too. Then I figured out...In the past (there were a lot of non-stop flights), if I depart from Shanghai, there will be a lot of Chinese (in the same airplane), I can just follow them.}" T03's previous experiences suggested a reliable and convenient infrastructure for international travelers with little English (or other local languages) proficiency. However, due to the travel bans, most passengers had to go to a third country first. The infrastructure for T08 broke down. She improvised communication with a digital translator to repair the breakdown.

\subsubsection{\textbf{Reducing Complexity in Stopovers}}

Another type of seams of the itinerary occurred in travelers' stopovers between each segment of the itinerary. For instance, participants who chose to "whitewash" themselves (8 out of 22) needed to stay in a third country for at least 14 days before they came to the US. As mentioned in prior work, in crises, stopovers in the trips were complex due to the unfamiliar local policies, cultural factors, etc \cite{Unger2021}. Therefore, reducing the complexity of the stopovers was also important to make the journey seamless. Overall, participants emphasized the quality of local tourism services. T05, a senior traveler, shared how she found a convenient hotel in Cambodia: "\textit{They (friends from WeChat groups) recommended me a hotel in Cambodia, they said the hotel is run by Chinese... including meals, all Chinese cuisine. Because I needed to stay in another country and I was worried about the language barriers. And then I went to the hotel, it felt like all passengers in the hotel was Chinese, and they all had the same destination (the US). I was very comfortable, we exchanged information about how to go to the US, what to bring for the trip, and what needs to be paid attention to.}" As revealed by T05, the hotel run by Chinese owners provided a place for Chinese travelers with similar backgrounds and purposes. Therefore, the hotel also served as a physical platform for social interaction and information exchange.  

\subsection{Handling Risks due to the COVID-19 Pandemic}
The third salient theme is about the risks related to covid. The coronavirus was one of the biggest safety concerns of travelers when they decided to go to the US, where 54 million cases had been identified and 823 thousand people died of the virus by the end of 2021 \cite{TheNewYorkTimes2022}. Due to the pandemic, travelers' trips were embedded with the risks of being infected or being rejected by public transportation due to infection. We present participants' work on handling covid-related risks in this section.  

\subsubsection{\textbf{Avoiding COVID Infection}}
Travelers needed to be careful about protecting themselves against the coronavirus in the pandemic. All participants mentioned how they got themselves protected on their trips, especially with the information they got from various information channels. Some of them talked about how they suited up for the trips. T13 demonstrated how she gathered the information for protection: "\textit{I searched for stories shared by people who came more recently on social media. I planned to go in August, so I look for people who traveled in June or July, and check what protections I should take. Also to what extent they protected themselves, like did they eat or drink during the journey. One single source cannot provide enough information. I got the information from many channels, like AirTicketNA, or the Chinese Student Association at my university. The association provides travel instructions for new coming students every year.}" The big number of questions showed that travelers' information needs were situated, and there was no single answer that could address all of them. Therefore, T13 took advantage of multiple channels, including not only professional sites but also personal networks, social media, and institutional support. 

In addition to personal equipment, interviewees also mentioned that they were concerned about the discrepancy in response to covid across different places. The discrepancy also influenced travelers' practices in protecting themselves against the virus. Like T14 said: "\textit{I took connecting flights, and the two stops were Hong Kong and Dubai. I had a concern about the covid policies in the two stops, like whether I would get delayed if there are cases nearby. I checked 1point3acres.com and knew some WeChat groups. Then I knew transferring was not a problem. (When I use 1point3acres.com)... They are online forums, they are not like instant messages, and they are more likely to have a delay in notification. So when I found people on the same flight, I just messaged them and get their WeChat contact or WeChat groups. (The online forums)...They are more like stepping stones to me, for example, if I find a WeChat group from the forum, I probably won't check it anymore and just used WeChat.}" T14's approach of switching across different ICTs showed how travelers utilized currently available sources to create an informative and convenient infrastructure for travel information. 

\subsubsection{\textbf{Managing COVID Test Reports}} 

In addition to avoiding infection, travelers also were concerned about their covid test reports. Getting covid tests before long-distance travel became a standard requirement after the breakout of the coronavirus. Local authorities needed to make sure that travelers will not be the carriers of the virus and that their trips will not put other passengers or residents in danger, and getting tested also helps travelers better monitor their health status. However, according to our participants (22 out of 22), who were international travelers coming from China to the US, managing test reports was a challenging task for them. 

Some participants pointed out the difficulty in finding reliable test centers. Like T19 said: "\textit{One thing is that the report should be in 24 hours. Preferably, the report needs to come out in 6-8 hours because I don't have so much time to wait. Then I noticed that some test institutions, they intentionally played it up, saying normal test reports come out after 24 hours, and expedited ones come out in only 4-6 hours. The normal ones are 40 CNY and the expedited ones are 600 CNY. But then I searched on Xiaohongshu, and some users recommended a couple of test centers that offer expedited reports, and the prices are the same as normal ones. Information like this, a lot of people if they don't know, they would be scammed by those institutions.}" T19's experience suggested that in crisis settings, trustworthy and reliable services were rare, and looking for such services required competencies like being tech-savvy.  

\subsection{Summary and Reflections}
In summary, travelers faced a large variety of breakdowns in the multifacet crisis. Specifically, the breakdowns manifested as malfunctions of the travel authorities, tourism industries, as well as public health sectors. The scale of the breakdowns was huge, affecting the global travel network. The breakdowns together lead to the breakdown of the international travel infrastructure. In the face of the travel infrastructure breakdown, travelers had to carry out their own work to fix it. While we briefly introduced travelers' countermeasures in this section, we will present a more detailed case study that focuses on the ICT use of travelers in the crisis in the next section.

\section{Case Study: Two Traveler Groups' ICT-Supported "Whitewashing"}
In this section, we highlight two cases that exemplify how travelers leveraged Information and Communication Technology (ICT) to navigate challenges during the crisis. Notably, the phenomenon of 'whitewashing' emerges as a significant countermeasure to the COVID travel ban imposed on Chinese travelers. Under the stipulations of this ban, individuals who had visited specific countries, including China, within the prior 14 days were prohibited from entering the US. Consequently, many travelers adopted a strategy of transiting through a third country, spending two weeks there, effectively "resetting" their travel history before proceeding to the US.

We specifically focus on these 'whitewashing' experiences because they present richer, more complex interactions. Such travelers, embarking on a journey that spanned at least three countries and extended beyond two weeks, were compelled to give added consideration to factors like the veracity of their travel details, personal security, and the dependability of their chosen routes. These narratives provide invaluable insights, with potential implications for the design of more resilient and adaptive travel infrastructures.

As reported by participants, eight participants chose the whitewashing approach (T01-08). Four of them were college students and all under 25 years old, and four of them were all above 45 years old and had little foreign language proficiency. The use of ICTs and experiences of whitewashing were different due to their different backgrounds. We closely examine each group's experiences in this section.

\subsection{Senior Travelers' Whitewashing}

Due to their language proficiency and technology literacy, when senior travelers used ICTs for whitewashing, they relied on the social network behind the ICTs more than technological affordance. According to participants, they preferred information channels that they were familiar with, such as WeChat. WeChat has been widely used among Chinese people all over the world. Especially, it is also an important information source for older adults. Specifically, participants' favorite function of WeChat was group chat. Like most group chats, WeChat Group allows users to add strangers to a group so that people with similar interests can discuss with each other (See \ref{fig:ICT}, a-b). 

The most important benefit of WeChat Group was information exchange. Traveling across multiple countries involves dealing with various departments and handling tremendous paperwork. WeChat Groups provided a space for participants to share key information, like T01 said: "\textit{In the group chat, there was a shared spreadsheet. Someone recorded, like where you can get English covid-test reports and when to do it, we just followed their instructions. And there was also a madam said we need to make sure that the centers provided 'fresh seal'. The first time I heard about the name, I was like, "what is a fresh seal"? And then I read all the chat history and figured it out. Because there would be other people asking the same questions, so you just read the chats and you'll figure out all the jargon.}" T01's experience suggests that WeChat Groups reduced cognitive loads so that senior participants could learn about their journey more easily.

Another significant role WeChat Groups played was to help participants find a reliable social network in the journey. Like T03 said: "\textit{There was a group chat, like specifically for grandparents, in the group chat, one guy added me and showed me the hotel. Later I found that the hotel was run by Chinese, and all people living there were grandparents or international students (heading to the US), it was very convenient because I don't speak English.}" Senior travelers also needed to think about their safety; traveling with others would make the entire journey safer. Using WeChat Groups, some travelers also found travel buddies from group chats. T02 said: "\textit{You also need to pay attention to your own safety, I heard a lot about robbery stories in Cambodia. Before we departed, people on the same flight, we created another smaller group chat specifically for us. In the small group chat, there were two students, they let me take care of their mothers during the journey.}" That said, relying on social support from WeChat Groups could also bring inconvenience. For instance, T01 got a travel buddy from WeChat Groups and they agreed to come to the US together. However, the travel buddy's schedule changed due to an emergency; T01 had to delay her plan accordingly to accommodate the travel buddy. 

Finally, all the senior participants selected Cambodia as the third country. According to the information from WeChat Groups, most of the Chinese travelers taking whitewashing went there, and the local tourism industry was Chinese-friendly. This option had the most reliable social network, or human infrastructure, for senior travelers.

\subsection{Student Travelers' Whitewashing}
Another group of travelers was student travelers. Compared to senior travelers, students were more familiar with ICTs and therefore had more options. In addition, they had better foreign language proficiency. Therefore, they could utilize more of the technological affordance of various ICTs than senior travelers. One of the most significant differences was the difference in terms of using WeChat Groups. While most of the senior participants used WeChat Groups to just get information, student travelers also shared their advice in WeChat Groups. In other words, they did not only use WeChat Groups as consumers, but they also used it as contributors.

While student travelers also used WeChat Groups, they also got information from other channels. Sometimes it was because of the different travel plans, like T04, a student who had to come back to the US in 15 days to get herself enrolled for a new semester, said: "\textit{Cambobia was the most popular choice in the WeChat Groups. But (In Cambodia) you need to make a deposit in a Cambodian bank, like a couple of thousand dollars, they hold the money for up to 17 days...I didn't have the time for the third country... I needed to go to a place with fewer procedures... I searched 1point3acres.com... So, I went to Mexico (which was less popular but required less time).}" T04 had to make a plan for her tight schedule, so she had to seek information from other sources other than WeChat Groups, like 1point3acres, a website specifically for Chinese residents or students in North America. 

Therefore, searching for information from various sources also required travelers' information literacy. One of the most important techniques was to find and reference successful precedents. T06 said: "\textit{Because the 'whitewashing' is a complex procedure, you have to deal with policies in China, the US, and the third country, and you need to cope with so many airlines, hotels, and the pandemic situations in places (you are going to stay). It (“whitewashing”) needs a lot of information, and you get it from precedents, you need the previous travelers to try and even fail so that you know what you can do and what you can’t. The most important thing is “data point” previous ones' experiences.}"

In addition, student travelers also paid close attention to the updates of information sources and then made changes to their plans. For instance, as mentioned in the previous section, T08 heard from a WeChat Group that one student's VISA was revoked during the transition in Tokyo: "\textit{After the rejection, the airline refused to provide any assistance to her. If she wanted to go back home, she had to go to another airport in Tokyo, but at that time, Japan did not allow any transfer passengers (without Japan VISA) due to the covid policies. So she had to buy tickets to some countries (that allow travelers without VISA), like Mexico... When we heard about this, we also ordered from the same airline. We canceled our tickets and bought another one. The new one provided very good service, if you get rejected by US customs, they take the responsibility to take you back to Singapore.}"

\subsection{Summary and Reflections}

To summarize, the two cases both presented how senior and student travelers deal with extreme uncertainties when traveling to a third country in a crisis. Both groups faced travel infrastructure breakdowns and had to conduct infrastructuring work with the support of ICTs. Senior travelers relied on tools they were familiar with and easy to use, like WeChat Groups; they primarily used the tools to connect to people so that they could form a reliable social network, such as a group of travel buddies taking care of each other in the journey. This option made senior travelers' trips much easier, but also made them depend on other travelers' information or schedule. However, student travelers had better foreign language proficiency and technology literacy. This allowed them to make more personalized and flexible travel plans. But this also required much more work in terms of searching, synthesizing, and sharing information.

\section{Discussion}

\subsection{Rethinking Travel Infrastructure: A Crisis Informatics Point of View}
Travel is a vulnerable activity, it involves handling people and policies in different places in an oftentimes limited period of time. The situation becomes even worse when crises happen, posing extreme uncertainty, anxiety, and even danger to travelers. From participants' practices in overcoming obstacles in the crisis, we perceived travelers' various needs, like making sense of travel policies, planning for trips, and handling covid risks. The various needs pointed out that, in a multi-facet crisis, ICT-mediated social interactions became one of the most important resources that travelers' relied on. Therefore, the needs also revealed a travel infrastructure, which was supposed to support travelers' various activities. In this section, we unpack the travel infrastructure and infrastructuring work, especially from a crisis informatics perspective.

The first vital component of travel infrastructure, as revealed by participants, is ICT infrastructure, which primarily consists of different social media platforms. In previous work, travel infrastructure usually referred to the physical substrate that supports traveling  \cite{10.1145/2063231.2063235}. In this study, travelers paid more attention to the information side infrastructure. This points out a unique weakness of travel infrastructure that, to travelers, the lack of information is as serious as the physical damage to transportation networks. In a travel infrastructure, there should be clear, transparent, and instructional information. However, such infrastructure did not exist due to the crisis. Therefore, travelers had to rebuild their own information infrastructure. We observe a specific pattern of leveraging the social media-centered ICT infrastructure. Because of the absence of official sources, travelers gathered together and discussed how to overcome the difficulties. Therefore, information was produced, shared, and exchanged among travelers. They also created or joined more specific information channels, like small group chats, to get more particular information. In the process, ICT infrastructure played an important role. First, it had various social media platforms. Travelers were able to gather information from the social media platforms they preferred. Second, ICT infrastructure allowed travelers to shift between different communication platforms. After finding people who had similar backgrounds or purposes, travelers could create group chats for discussion. Such support cannot be provided by any single social media platform, it has to be an "\textit{information ecosystem} \cite{10.1145/2998181.2998290}" that covers diverse populations and offers flexible services.

The second key component of travel infrastructure is the underlying human infrastructure. Lee and colleagues \cite{Lee2006} coined "\textit{human infrastructure}" to denote "\textit{the arrangements of organizations and actors that must be brought into alignment in order for work to be accomplished.} \cite{Lee2006}" Travelers' work on overcoming various challenges was primarily supported by the ICT infrastructure, where numerous travelers exchanged information. The ICT infrastructure would not exist without the travelers who produced and shared such information. Therefore, the travelers together formed a human infrastructure that supported the travel infrastructure, too. Reciprocity played an important role in enhancing the relationships among travelers. Travelers asked questions in the groups, and they also provided answers to other travelers. The selfless information exchange fostered a sense of mutual help and further formed a community. To be noted, the selfless information exchange occurred not only among travelers who successfully came to the US but also among travelers who failed to do so. Thanks to such failed precedents' experiences, later travelers were able to update and modify their plans accordingly so that they would not make the same mistakes. In this sense, the human infrastructure significantly increased travelers' possibility of successfully finishing the journey.  

\subsection{Personal Infrastructuring: The ICT-Supported Grassroots Work in a Multifacet Crisis}

In addition to the travel infrastructure, we also demonstrate travelers' work that fixed the breakdowns. We turn to personal infrastructuring to comprehend travelers' efforts in their journeys. Personal infrastructuring, based on Rajapakse et al.'s definition \cite{10.1145/3196709.3196749}, refers to the work "\textit{of learning about how to navigate the world, what support is available, and how to obtain and design new support through various organisational infrastructures} \cite{10.1145/3196709.3196749}." Travelers had diverse backgrounds like age, language proficiency, schedule, etc. The diversity of travelers, despite the same goal, showed that there would not be a single solution to all travelers' problems. Therefore, they had to conduct personal infrastructuring to "\textit{devise a personal approach to existing infrastructure} \cite{10.1145/3196709.3196749}" when "\textit{standard technologies and protocols do not work for them} \cite{10.1145/3196709.3196749}." Our study aligns with the key notion of personal infrastructuring. For example, Rajapakse emphasized the importance of "\textit{boundary objects}" in personal infrastructuring, which was proposed by Star and Griesemer \cite{Star1989}, arguing that the boundary object should have three qualities 1) boundary object "\textit{resides between social worlds} \cite{10.1145/3196709.3196749}," 2) boundary object should be "\textit{worked upon by local groups who maintain its identity as a common object, while making it more specific} \cite{10.1145/3196709.3196749}," and 3) "\textit{groups that cooperate without consensus refer back-and-forth to both vague and specific forms of the object} \cite{10.1145/3196709.3196749}." As elaborated in previous sections, social media, especially WeChat groups, played the role of connecting travelers with different social backgrounds and allowing back and forth communication in terms of solving travelers' confusion and questions.

That said, our findings also provide rich implications that extend the concept of personal infrastructuring and connect it to previous infrastructure literature in more aspects. First, our findings contextualize the personal infrastructuring work by specifying travelers' "\textit{torqued}" experiences. Bowker and Star \cite{Bowker1999} used "\textit{torque}" to refer to experiences of infrastructure users when "\textit{the ‘time’ of the body and of multiple identities cannot be aligned with the ‘time’ of the classification system} \cite{Bowker1999}." Moitra et al. \cite{Moitra2021} considered it as "\textit{the lived experience of excluded users of an infrastructure-in-use} \cite{Moitra2021}." In the current project, travelers were torqued because they were traveling to a foreign country. On the trip, they neither had any privileges nor official guidelines. At the travel policy level, they were torqued because the travel infrastructure excluded them. They had to deal with the torque to get approval from the travel infrastructure in the US, and the approach was to conduct personal infrastructure. Due to the exclusion from official sources, travelers had to come up with solutions to meet their various needs. Emphasizing the torqued experience of travelers provided a more specific context of and reason for travelers' situations, problems, and their personal infrastructuring work.

We also foreground the "\textit{infrastructural competence}" in some participants' personal infrastructuring work. According to Erickson and Jarrahi \cite{10.1145/2818048.2820015}, having "\textit{infrastructural competence}" means "\textit{to be able to recognize where infrastructural seams may have generative, rather than exclusionary, properties and then to draw upon this sociotechnical insight to fashion and implement an infrastructural strategy to achieve a desired goal.} \cite{10.1145/2818048.2820015}" We have shown the importance of the experiences shared by previous travelers; that said, it does not mean that participants just imitated such experiences. Many participants, like T08, emphasized the importance of reasoning based on shared information, echoing previous work on comprehending information in crises \cite{10.1145/3025453.3025891}. Such emphasis suggested travelers' "\textit{infrastructural competence}" while fixing the broken infrastructure.

\section {Design Implications}
After discussing the travel infrastructure and the infrastructuring work conducted by the travelers, we propose design implications for a more resilient travel infrastructure in crisis. The implications pivot around designing more responsive ICT infrastructure and more robust human infrastructure as well as supporting travelers' infrastructuring work. Although travel infrastructure involves various stakeholders like authorities, agencies, and travelers, our implications are primarily for authorities and agencies who are able to design, build, and manage a large-scale travel infrastructure.

The first design implication is about allowing travelers to communicate with each other through official channels. In the crisis, most authorities lack a timely and accurate understanding of international travel. Only travelers themselves have the first-hand information that is useful. Therefore, if the official channels are not working, getting help from travelers could be an important approach. Airlines or travel authorities can provide discussion sections on their websites or create official WeChat Groups to allow travelers to leave questions and answers. The officials can also utilize their accounts on social media platforms to encourage communication among travelers. This provides travelers with both convenient and secure venues to exchange information. For the discussion section or group chats, there should be 1) easy access to join so that travelers can find and participate in the discussion; 2) share buttons so that travelers can share the chats with others; 3) functions all personalized discussion sections or group chats for travelers with more specific requirements. Therefore, the official communication channels could form a responsive ICT infrastructure but also a reliable human infrastructure.

The second implication is to encourage story-sharing among travelers. According to participants, travel stories were the most important information sources. Therefore, travel authorities can invite people to share stories, such as 1) travelers (while successful stories provided useful instructions to travelers, failed ones also helped them avoid certain situations); 2) local residents (they can share more accurate local covid status and travel policies). Travel authorities can also provide more technical support to encourage story-sharing among travelers. For instance, the document should record not only strategies used in the trips but also other background information (itinerary, schedule, VISA status, English proficiency). Therefore, stories can be stored in a spreadsheet, based on their background so that other travelers can easily retrieve and share them. Officials can also review and organize the stories, then make a document with an organized table that shows links to different travel stories, so that travelers with little background knowledge can also understand them. 

\section {Limitations and Future Work}



To investigate travelers' experiences, we conducted an interview study. We focused on travelers who had successfully finished their journeys to the US. That said, as revealed by participants, travelers who failed to come to the US also had rich interactions in terms of using ICTs to try to address their problems. In addition, some failed travelers, like the female traveler mentioned by T08, encountered more complex and urgent situations. Therefore, despite their failures, those travelers' experiences can also provide rich implications for crisis informatics and HCI. In this regard, future research can be done on a more diverse traveler population including not only successful but also failed travelers.

Besides, the story shared by T19, who encountered test centers charging unreasonably high prices for covid test reports, showed that scamming could be a serious issue among travelers. That said, such issues only took place between organizations and travelers. Given travelers' heavy use of social media, where most interactions were taken place among strangers, scamming among travelers could also happen. Prior work on collaboration in crisis pointed out the problems. For instance, Semaan \cite{10.1145/2063231.2063235} mentioned that residents in a war were afraid of traveling because they could be ransomed by strange taxi drivers. Therefore, further work can be done to explore issues like privacy and trust among travelers, who shared so much personal information (e.g. flight, hotel) and collaborate with so many strangers with various backgrounds.    

\section {Conclusion}
International travel involves extreme uncertainty, especially in crises. In the current study, we report an investigation on Chinese travelers' encounters in their journeys to the US in a multi-facet crisis and their use of ICTs to overcome difficulties in the journeys. We interviewed 22 Chinese travelers who had successfully come to the US in the crisis. We found three themes: overcoming policy restrictions, stitching itinerary seams, and handling covid risks. The work has rich implications for the literature on crisis informatics, infrastructure, and tourism.

\begin{acks}

\end{acks}

\bibliographystyle{ACM-Reference-Format}
\bibliography{Travel Study}

\appendix

\end{document}